\documentclass[12pt]{iopart}
\expandafter\let\csname equation*\endcsname\relax
\expandafter\let\csname endequation*\endcsname\relax
\usepackage{graphicx}
\usepackage{siunitx}
\usepackage[version=4]{mhchem}
\usepackage{hyperref} 
\usepackage[capitalize]{cleveref}
\usepackage[square,numbers]{natbib}
\usepackage{iopams}
\usepackage{caption}
\usepackage{subcaption}
\usepackage{multirow}
\usepackage{booktabs}
\usepackage[title]{appendix}

\begin{document}

\title[Importance of Configurational Charge Ordering for MLIPs]{Importance of Configurational Charge Ordering for Machine Learning Interatomic Potentials}

\author{Martin Hoffmann Petersen$^{1,2,3}$*, Steen Lysgaard$^1$, Arghya Bhowmik$^1$, Kedar Hippalgaonkar$^{2,3}$, and Juan Maria Garcia Lastra$^1$*} %

\address{1 Technical University of Denmark, Department of Energy Conversion and Storage, Lyngby, 2800, Denmark}
\address{2 School of Materials Science and Engineering, Nanyang Technological University, Singapore 639798}
\address{3 Berkeley Education Alliance for Research in Singapore, CREATE Tower, Singapore 138602 }

\ead{martin.petersen@ntu.edu.sg, jmgla@dtu.dk}

\begin{abstract}
Machine learning interatomic potentials (MLIPs) enable large-scale atomistic simulations but remain challenged in describing mixed-valence materials where charge ordering strongly influences thermodynamic stability. Here we investigate the role of configurational charge ordering in MLIP structural optimization of the battery cathode material \ce{NaFePO4}. We show that conventional MLIPs fail to reproduce the correct stability of intermediate \ce{Na} concentrations because structural optimization leads to incorrect \ce{Fe^{2+}}/\ce{Fe^{3+}} charge assignments, resulting in incorrect energy ordering and convex-hull predictions. Analysis of the \ce{Fe^{2+}}/\ce{Fe^{3+}} charge assignment during structural optimization reveals that MLIPs are unable to capture the configurational charge ordering, leading to energy errors that make it difficult to identify the experimentally observed phase as the true ground-state structure. To address this limitation, we introduce an approach that encodes charge-state information directly into the MLIP representation by distinguishing between \ce{Fe^{2+}} and \ce{Fe^{3+}} environments during training. Retraining CHGNet, MACE and MACELES with this representation enables accurate structural optimization, correct identification of charge ordering, and improved agreement with density functional theory convex hulls. Our results demonstrate that incorporating configurational charge ordering into the MLIP representation is essential for modeling charge-disordered materials. Beyond recovering the correct ground state, this encoding allows a specific charge ordering to be imposed and held fixed during optimization, a capability unconstrained DFT struggles to guarantee, and we show that it generalizes across four polyanion frameworks and four transition-metal ions (Fe, Mn, Co, Ni), establishing a broadly applicable strategy for modeling mixed-valence transition-metal systems.
\end{abstract}

\section{Introduction}
Computational materials discovery has long been an integral component of scientific research, playing a central role in interpreting and complementing experimental studies while also enabling the prediction of novel materials and guiding experimental design \cite{MannodiKanakkithodi2021}. For several decades, first-principles methods such as Density Functional Theory (DFT) have served as the cornerstone of computational materials science, providing reliable access to electronic, structural, and thermodynamic properties \cite{DFT_book,Emery2017,devi2022effect}. More recently, Machine-Learning Interatomic Potentials (MLIPs) have emerged as an efficient and accurate alternative to DFT for many applications, enabling simulations at length and time scales that would otherwise be computationally prohibitive and facilitating large-scale exploration of complex materials spaces \cite{yang2024mattersim,unke2021machine,Kulichenko2024}.

Despite these advances, conventional MLIPs remain inherently limited by the information encoded in their input representations, which are typically based solely on atomic coordinates and chemical species and therefore lack an explicit description of electronic degrees of freedom \cite{Bonneau2025,Oudah2025,Zhou2006}. This omission can lead to systematic errors in systems where identical atomic species occur in multiple charge states, as the local atomic environment alone may be insufficient to uniquely determine the underlying electronic configuration and associated interactions\cite{unke2019physnet,Bartk2013}. Morever, recent generative models for inorganic materials have similarly found that structure-only representations (atom types, coordinates, lattice parameters) are insufficient, and that adding charge density as an explicit modality improves both generation quality and inverse-design performance \cite{Park2026}, suggesting this is a general limitation of ML approaches to materials, not one specific to MLIP optimization.

These limitations are particularly critical for battery cathode materials, where electrochemical performance is intrinsically linked to redox activity and the presence of multiple charge states of atomic species \cite{Kitchaev2018,Whittingham2004,chen,Simon,de2023nanosecond}. Neglecting explicit electronic information can therefore compromise the accurate description of transition-metal chemistry, defect formation, and structural disorder, ultimately limiting the predictive capability of MLIPs.

\begin{figure}
    \centering
    \includegraphics[width=1.0\textwidth]{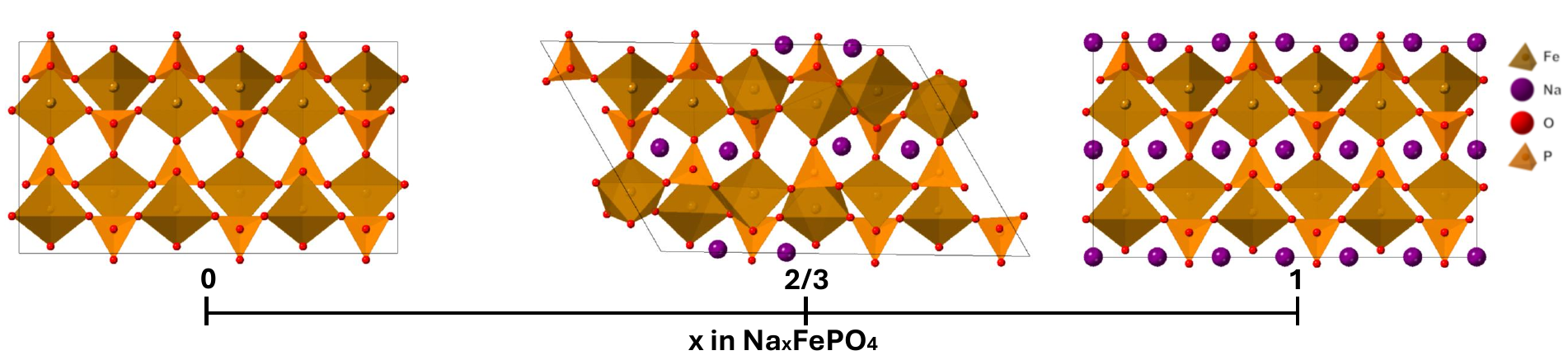}
    \caption{The stable experimental crystal structures of \ce{NaFePO4} at different concentration of sodium \cite{kyoto,kyoto_cif}.}
    \label{fig:crystal}
\end{figure}

To investigate these challenges, we focus on the olivine cathode material \ce{NaFePO4}, which requires consideration of multiple charge states associated with different sodium concentrations, ranging from the fully discharged state, \ce{NaFePO4}, to the fully charged state, \ce{FePO4}. Across this compositional range, iron acts as the redox-active center and undergoes a change in oxidation state from \ce{Fe^{2+}} to \ce{Fe^{3+}} to compensate for the removal of \ce{Na^+} ions. Furthermore, \ce{NaFePO4} is known to exhibit a phase transition at approximately 66\% sodium content during (de)sodiation \cite{kyoto_cif,kyoto}, making it an ideal benchmark system for assessing whether MLIPs can accurately capture distinct thermodynamic phases.

By constructing the composition–energy convex hull using the fully charged and fully discharged states as reference points, the thermodynamically stable configurations occurring during charge and discharge can be determined. This provides a stringent test of a model’s ability to describe charge-state-dependent chemistry in complex battery materials. The experimentally observed stable configurations of \ce{NaFePO4} at different \ce{Na} concentrations are visualized in \cref{fig:crystal}, including the reference states and the stable monoclinic phase at 66\% sodium concentration \cite{kyoto_cif,kyoto}.

The olivine cathode material \ce{NaFePO4} contains distinct Wyckoff sites \cite{Souvignier2016,ai4science2023crystal,Simon,Ismail2024} for \ce{Na}, \ce{Fe}, and \ce{P}, as well as three Wyckoff sites for \ce{O}. The Wyckoff sites associated with \ce{Na} can be partially occupied by \ce{Na} ions and vacancies, giving rise to \emph{configurational spatial ordering} as \ce{Na} ions distribute across the available symmetry-equivalent sites. In contrast, the \ce{Fe} Wyckoff sites remain fully occupied, but the Fe atoms can exist as either \ce{Fe^{2+}} or \ce{Fe^{3+}} depending on the sodium concentration. The different spatial arrangements of these charge states introduce an additional degree of freedom that contributes to the free energy of the system. Following Ref.~\cite{Zhou2006}, we refer to this contribution as \emph{configurational charge ordering}.

To account for the configurational spatial ordering arising from \ce{Na}–vacancy disorder, a Genetic Algorithm (GA) \cite{Jennings2019,Lysgaard2013} is employed to efficiently sample low-energy configurations at each sodium concentration in \ce{NaFePO4}. The GA optimizes the distribution of \ce{Na} ions over the available Wyckoff sites using energies predicted by the MLIPs, enabling the identification of energetically favorable atomic arrangements.

\begin{figure}
    \centering
    \includegraphics[width=1.0\textwidth]{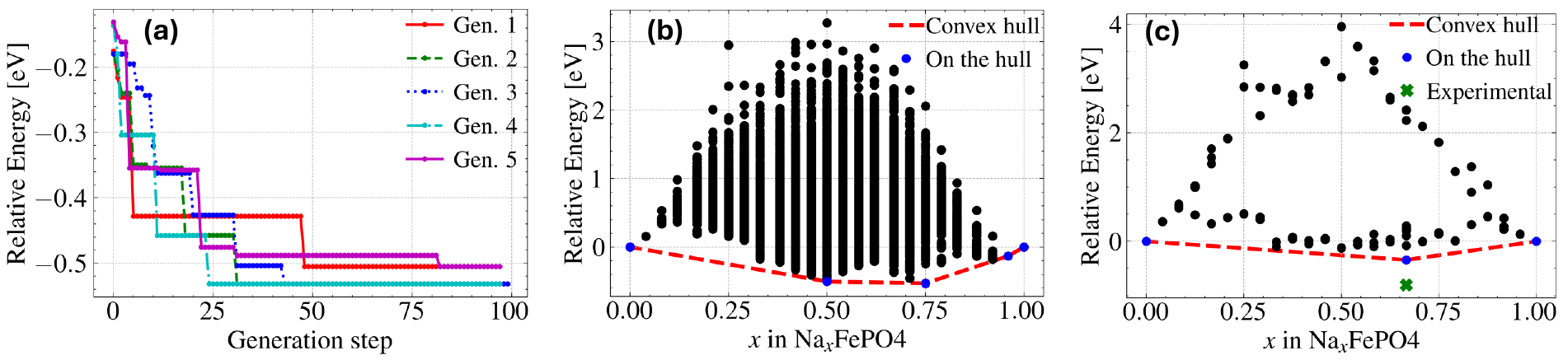}
    \caption{a) the different GA generations and their steps. b) The total MACE MLIP driven GA convex hull based on all the GA generations c) DFT optimized structures based on the MACE MLIP driven GA convex hull }
    \label{fig:GA_result}
\end{figure}

Using a trained MACE model \cite{batatia2022mace} on the polyanion sodium cathode materials dataset \cite{Polyanion_dataset}, which includes several cathode materials exhibiting Na–vacancy disorder, we performed an MLIP-driven GA optimization to explore the configurational space of \ce{NaFePO4}. For each candidate configuration, the MLIP first performs a structural optimization starting from the symmetrized crystallographic unit cell, in which the Wyckoff-site occupations are explicitly defined. Throughout this work, this initial structure is referred to as the \emph{pre-optimized} structure, while the structure obtained after structural optimization is referred to as the \emph{optimized} structure.

As visualized in \cref{fig:GA_result}b, the MLIP-driven convex hull predicts five stable phases instead of the three observed experimentally and fails to capture the experimentally verified configuration at 66\% sodium concentration. To further analyze this discrepancy, the three lowest-energy and three highest-energy structures identified by the MLIP-driven GA were selected for DFT optimization. In addition, seven low-energy configurations at 66\% Na concentration were included, yielding a total of ten low-energy candidate structures for further evaluation. The resulting DFT-optimized convex hull is shown in \cref{fig:GA_result}c, consisting of 146 structures. Comparison with the MLIP-driven GA convex hull reveals substantial deviations, indicating that the MLIP fails to accurately capture the thermodynamic stability across the sodium composition range. 

In this work, we investigate the origin of the discrepancy between the DFT-optimized convex hull and the MLIP-optimized convex hull and examine its relation to configurational charge ordering, arising from different spatial arrangements of \ce{Fe^{2+}} and \ce{Fe^{3+}} ions. As this study focuses on a Fe-based cathode material, the polyanion sodium cathode materials dataset \cite{Polyanion_dataset} was filtered to include only Fe-containing compounds, for which a clear distinction between \ce{Fe^{2+}} and \ce{Fe^{3+}} can be made using the atomic magnetic moment as a proxy for the oxidation state. Using this dataset, we retrain both MACE and the recently proposed MACELES model \cite{Kim2026} to evaluate the effect of explicitly incorporating charge-state information and long-range electrostatic interactions.
To further assess the impact of electronic information on MLIP performance, we additionally employ the CHGNet \cite{deng2023chgnet} model in this work. CHGNet predicts atomic magnetic moments \cite{mulliken1955electronic} by extracting embedded information from the second-to-last layer of the message-passing neural network, providing an implicit description of the local electronic environment. 

Our results demonstrate the importance of incorporating configurational charge ordering information directly into MLIP training when charge disorder is present in a material. To achieve this, we introduce a \emph{charge-ordering-encoding} of the MLIP, where a surrogate-species approach is utilized in which \ce{Fe^{3+}} ions are temporarily replaced by \ce{Ga^{3+}} during structural optimization and training-data construction, exploiting the fact that \ce{Ga} preferentially adopts a stable +3 oxidation state and has an ionic radius similar to that of \ce{Fe} \cite{Ga_method,Zhou2006}. This allows the MLIP representation to explicitly distinguish \ce{Fe^{2+}} from \ce{Fe^{3+}} environments during training. This encoding also allows a specific \ce{Fe^{2+}}/\ce{Fe^{3+}} arrangement to be imposed and held fixed throughout structural optimization, a capability that unconstrained DFT struggles to guarantee. This makes the approach useful not only for recovering the correct ground state, but also as a pre-optimizer for isolating the energetic consequences of a prescribed charge ordering. We establish \ce{NaFePO4} as a stringent and physically motivated benchmark system for evaluating the capability of MLIPs to capture charge-state-dependent chemistry, atomic disorder, and phase stability in battery cathode materials. While \ce{NaFePO4} serves as our primary benchmark, the approach is not limited to it or to a single transition-metal system. This is verified by applying the same charge-ordering-encoding approach to a generalized MLIP trained across four polyanion cathode frameworks and four transition-metal ions (Fe, Mn, Co, Ni), and evaluating its performance on out-of-distribution structures containing multiple co-doped transition-metal species.

\section{Results}
\subsection{Importance of configurational charge ordering in structural optimization}
From the MLIP-driven GA, the ten lowest-energy structures for the \ce{Na_{0.66}FePO4} composition were examined, and none of them reproduced the Na ordering observed in the experimentally verified phase \cite{kyoto_cif}. By utilizing the Na ordering of the experimentally verified monoclinic phase in \cref{fig:crystal}, we were able to map it onto an orthorhombic structure resembling the crystal lattice of the fully charged and fully discharged states of \ce{NaFePO4}. This mapped orthorhombic structure is referred to as the experimental structure for the remainder of this work.

\begin{figure}
    \centering
    \includegraphics[width=1.0\textwidth]{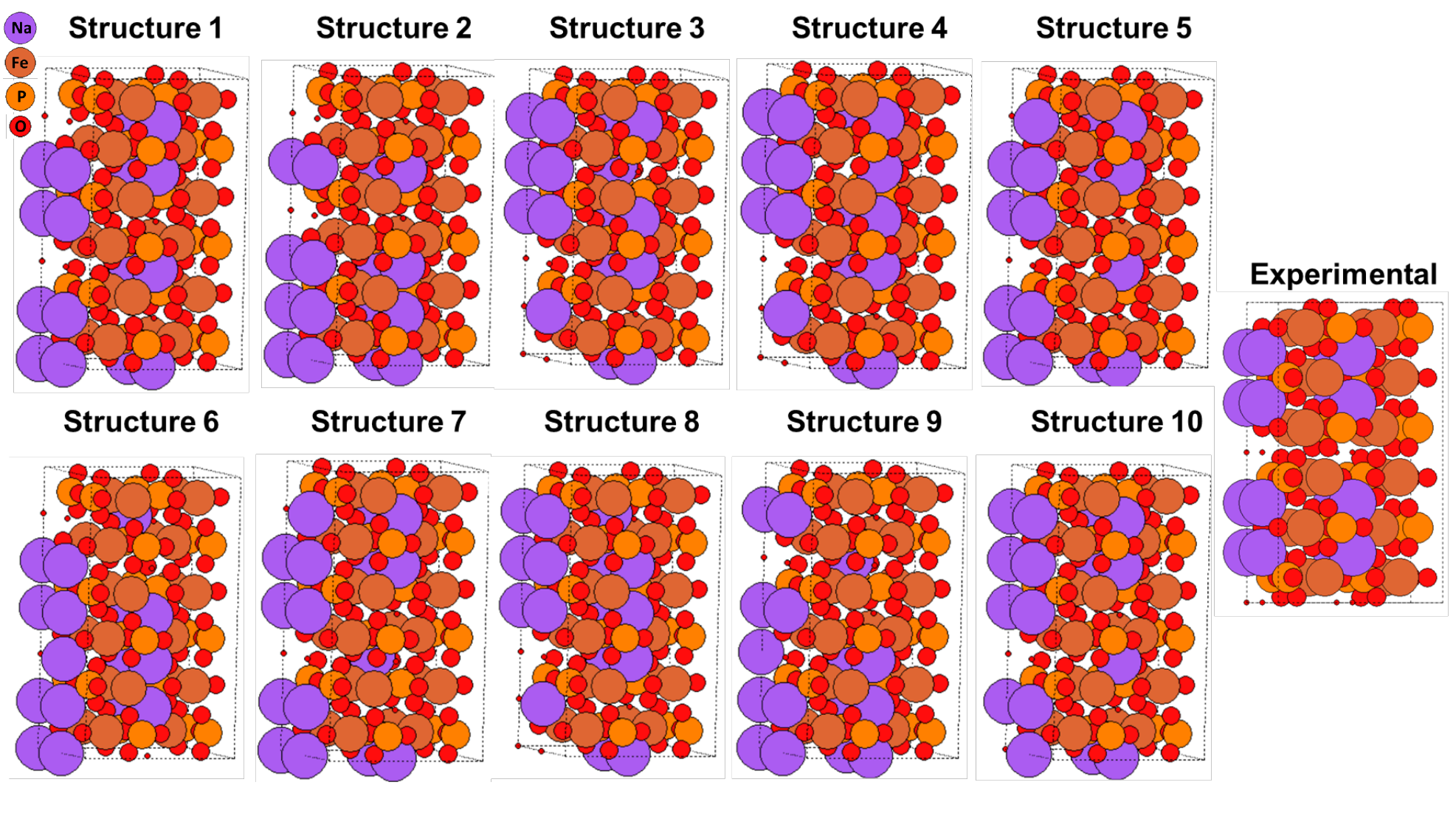}
    \caption{The ten lowest-energy structures identified by the MACE-driven GA at 66\% Na concentration, together with the experimentally verified phase used as a reference. Small red dots indicate Na vacancies 
    }
    \label{fig:GA_structures}
\end{figure}

The ten lowest-energy structures identified by the GA, together with the experimental structure, are visualized in \cref{fig:GA_structures}. These eleven structures form a unique benchmark set, with the objective of correctly identifying the experimentally verified structure as the lowest-energy configuration and to recreate the energetic ordering from DFT.

\begin{table*}[htb!]
\centering
    \caption{Relative energies [eV] of the ten lowest-energy structures identified by the MACE-driven GA at 66\% Na concentration, together with the experimentally verified phase used as a reference. Energies are reported for the three MLIPs and DFT and are given relative to the energy of the optimized experimental structure, yielding a zero reference point for each column.}
    \begin{tabular}{ccccc}
    \toprule
    & CHGNet & MACE &MACELES  & DFT \\
    \midrule
Structure 1&	0.683&	0.220&	0.532&	0.720\\
Structure 2&    0.921&	0.464&	\textbf{-0.086}&	0.457\\
Structure 3&    0.393&	0.262&	0.195&	1.000\\
Structure 4&    0.393&	0.262&	0.195&	0.996\\
Structure 5&    0.393&	0.262&	0.195&	0.902\\
Structure 6&    0.393&	0.262&	0.195&	1.003\\
Structure 7&    0.393&	0.262&	0.195&	0.996\\
Structure 8&    0.677&	\textbf{-0.067}&	0.187&	1.086\\
Structure 9&    0.677&	\textbf{-0.067}&	0.187&	1.049\\
Structure 10&   0.677&	\textbf{-0.067}&	0.187&	1.016\\
Experimental&   \textbf{0.000}&	0.000&	0.000&	\textbf{0.000}\\

        \bottomrule
    \end{tabular}
    \label{table:GA_structures}
\end{table*}

Optimizing these eleven structures using CHGNet, MACE, and MACELES resulted in the relative energies shown in \cref{table:GA_structures}, along with the corresponding DFT values. All energies are referenced to the experimentally derived structure, yielding a zero reference point for each MLIP and for DFT. MACE and MACELES fail to identify the experimentally verified phase as the lowest-energy structure, whereas CHGNet correctly predicts it as the ground state.

However, although CHGNet correctly identifies the experimental phase as the lowest in energy, the predicted ordering of the remaining structures deviates significantly from the DFT results. In particular, the structures that correspond to the second- and third-lowest energies in DFT (structures 2 and 1, respectively) are incorrectly ordered, and structure 2, the second lowest in energy according to DFT, is instead predicted to be the highest in energy. This behavior suggests that the inclusion of auxiliary electronic information, such as magnetic moments, may improve the description of mixed-valence systems where charge ordering plays an important role in determining thermodynamic stability.


MACELES exhibits a modest improvement over the baseline MACE model, identifying the second-lowest-energy structure as the predicted ground state, whereas MACE instead identifies one of the high-energy structures as the predicted ground state. Nevertheless, both models produce energy rankings that differ substantially from the DFT reference. This observation indicates that incorporating long-range electrostatic interactions alone is insufficient to recover the correct energetic ordering, suggesting that the primary source of error is not the treatment of electrostatics but rather the inability to accurately represent the underlying \ce{Fe^{2+}}/\ce{Fe^{3+}} charge-state distribution.

Configurational charge ordering plays an important role in charge-disordered materials. If the charge states are incorrectly assigned in a DFT calculation, the resulting charge density can lead to an energy that does not correspond to the true global minimum. To verify this hypothesis, we employ a previously established method \cite{Ga_method}, described in \cref{sec:benchmark_datasets}, that enables control over which specific \ce{Fe} ions adopt the \ce{Fe^{3+}} oxidation state upon desodiation.


\begin{figure}
    \centering
    \includegraphics[width=0.75\textwidth]{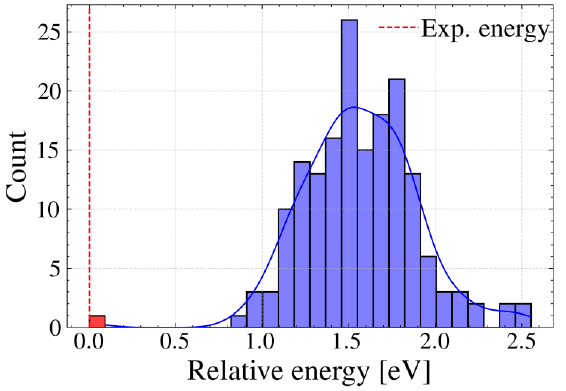}
    \caption{Distribution of DFT energies for 171 different \ce{Fe^{2+}}/\ce{Fe^{3+}} arrangements in the experimental \ce{Na_{0.66}FePO4} structure. The experimentally derived configuration is indicated with red and lies significantly below the mean of the sampled distribution.}
    \label{fig:electronic_entropy}
\end{figure}

To account for contribution of configurational charge ordering in the \ce{Na_{0.66}FePO4} experimental structure, different arrangements of \ce{Fe^{3+}} ions were considered. For each configuration, the procedure, described in \cref{sec:benchmark_datasets}, was applied to obtain the corresponding DFT energy after structural optimization with the prescribed charge ordering. As shown in \cref{fig:electronic_entropy}, the resulting energies, reported relative to the experimentally derived structure defining the zero reference point, follow an approximately Gaussian-like probability distribution with a mean of \SI{1.564}{eV} and a standard deviation of \SI{0.331}{eV}. This distribution clearly reflects the energetic penalty associated with less favorable charge-ordering configurations. Notably, the experimental structure lies significantly below the mean of the sampled distribution, indicating that the experimentally observed configuration corresponds to a particularly favorable arrangement of \ce{Fe^{2+}} and \ce{Fe^{3+}} ions within the lattice.

These results demonstrate that variations in the charge ordering alone can produce large energy differences, highlighting the importance of configurational charge ordering in determining the thermodynamic stability of mixed-valence cathode materials.

This raises the question of how MLIPs, and comparatively DFT, handle configurational charge ordering during structural optimization. CHGNet predicts atomic magnetic moments, providing a useful probe of the underlying charge distribution in the material. In particular, \ce{Fe^{2+}} typically exhibits a magnetic moment of approximately 4$\mu_B$, whereas \ce{Fe^{3+}} exhibits a magnetic moment of approximately 5$\mu_B$. This clear difference in magnetic moments between the two oxidation states enables the identification of their spatial distribution within the crystal lattice during structural optimization.

\begin{figure}
    \centering
    \includegraphics[width=0.8\textwidth]{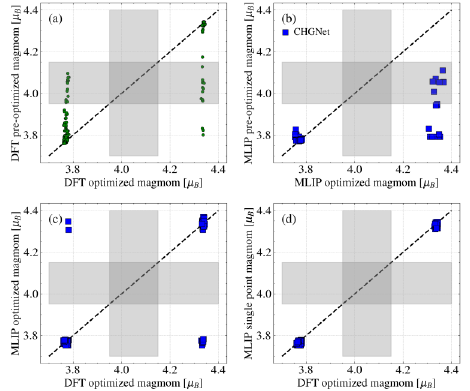}
    \caption{Comparison of \ce{Fe} magnetic moments predicted by DFT and CHGNet for the eleven structures shown in \cref{fig:GA_structures}. 
    (a) Magnetic moments of \ce{Fe} atoms in the initial DFT pre-optimized structures compared with those obtained after DFT structural optimization.  
    (b) Magnetic moments of the pre-optimized structures predicted by MLIP compared  with those obtained after structural optimization. 
    (c) Magnetic moments of the MLIP-optimized structures compared with those obtained from DFT optimization. 
    (d) Magnetic moments from single-point MLIP calculations on the DFT-optimized structures compared with those obtained from DFT optimized structure. For all plots, the grey region highlights the range where the oxidation state assignment is ambiguous.}
    \label{fig:MLIP_scatter}
\end{figure}

We first examine the role of configurational charge ordering during DFT structural optimization. In \cref{fig:MLIP_scatter}a), the magnetic moments of each \ce{Fe} atom in all eleven structures from \cref{fig:GA_structures} are compared between the initial pre-optimized structures and the corresponding DFT-optimized structures. We observe that the charge states of most \ce{Fe} atoms are already determined in the initial pre-optimized cell before the structural optimization begins. However, there exists a region with 10.6\% of the atoms, highlighted in grey, where the charge states of the \ce{Fe} atoms remain ambiguous. Although a few outliers are present (1.1\%), where \ce{Fe} atoms are initially assigned an incorrect charge state, this analysis indicates that the charge states of the \ce{Fe} atoms are effectively determined at the start of the DFT structural optimization. 

To assess how well CHGNet predicts the magnetic moments at this crucial stage of the optimization process, the magnetic moments of the pre-optimized structures predicted by the MLIP are compared with those predicted after MLIP structural optimization in \cref{fig:MLIP_scatter}b). It is evident that CHGNet assigns nearly all Fe atoms to the same charge state, which in this case corresponds to \ce{Fe^{2+}}. As a result, the \ce{Fe^{2+}} sites are correctly assigned by construction, whereas the \ce{Fe^{3+}} sites are either incorrectly assigned to the \ce{Fe^{2+}} state (13.3\%) or fall within the ambiguous region (12.1\%). This uncertainty in the charge states during the initial stage of the structural optimization may therefore influence the subsequent structural relaxation, potentially leading to local chemical environments with incorrectly assigned oxidation states.

This hypothesis is investigated further in \cref{fig:MLIP_scatter}c), where the magnetic moments of the DFT-optimized structures are compared with those of the MLIP-optimized structures for all eleven configurations. We immediately observe that CHGNet incorrectly assigns the oxidation states of \ce{Fe^{2+}} (1.8\%) and \ce{Fe^{3+}} (9.8\%) to wrong charge state. Such misassignments are likely to lead to incorrect energy predictions, as demonstrated by the configurational charge ordering analysis in \cref{fig:electronic_entropy}, where an incorrect charge ordering can result in large energy deviations. This behavior may therefore explain why the ordering of the lowest-energy structures in \cref{table:GA_structures} is incorrectly predicted by all MLIPs. This observation is further supported in \cref{app:charge_singlepoint}, where similar behavior is observed for the MACE and MACELES model.

The incorrect assignment of oxidation states may originate from the early stages of the structural optimization, where the MLIPs establish local chemical environments corresponding to local energy minima. This raises the question of whether the MLIPs can correctly identify the charge states of \ce{Fe} when a more energetically favorable local chemical environment is provided, or whether they are fundamentally unable to resolve the charge distribution within the material. To investigate this, the MLIPs are used to perform single-point calculations on the DFT-optimized structures in order to evaluate whether the correct charge-state-dependent energetics can be recovered.

This comparison is shown in \cref{fig:MLIP_scatter}d), where we observe that CHGNet correctly identifies the charge distribution of \ce{Fe}. This indicates that the MLIPs are able to correctly recognize the charge distribution of \ce{Fe} atoms when an appropriate local chemical environment is provided. 

These results demonstrate that configurational charge ordering plays a crucial role in determining the correct energy ordering of charge-disordered structures. Incorrect assignments of oxidation states during structural optimization can lead to less energetically favorable local chemical environments and consequently incorrect energy predictions. While the MLIPs are capable of identifying the correct charge distribution for a given local chemical environment, they remain unable to reliably recover the correct charge ordering when structural optimization begins from an initial ambiguous configuration. This limitation represents a significant challenge for the application of MLIPs to charge-disordered materials in computational simulations.

\subsection{Encoding configurational charge ordering directly into machine learning interatomic potentials}
Having established the importance of configurational charge ordering for structural optimization, we identified a key limitation of conventional MLIPs: they are unable to model and utilize configurational charge ordering during structural optimization. Consequently, MLIP-based optimizations can converge to local chemical environments that differ from those obtained with DFT and may correspond to less energetically favorable configurations.

\begin{table*}[htb!]
\centering
    \caption{ Relative energies [eV] of the ten lowest-energy structures identified by the MACE-driven GA at 66\% Na concentration, together with the experimentally verified phase used as a reference. Energies are reported for the three charge-ordering-encoded MLIPs and DFT and are given relative to the energy of the optimized experimental structure, yielding a zero
    reference point for each column.}  
    \begin{tabular}{ccccc}
    \toprule
    &  CHGNet & MACE & MACELES & DFT\\
    \midrule
Structure 1	&0.558	&0.438	&0.640	&0.720\\
Structure 2	&0.121	&0.249	&0.292	&0.457\\
Structure 3	&0.764	&0.609	&0.779	&1.000\\
Structure 4	&0.764	&0.609	&0.836	&0.996\\
Structure 5	&0.784	&0.628	&0.771	&0.902\\
Structure 6	&0.763	&0.609	&0.836	&1.003\\
Structure 7	&0.764	&0.609	&0.836	&0.996\\
Structure 8	&0.805	&0.714	&0.833	&1.086\\
Structure 9	&0.805	&0.714	&0.833	&1.049\\
Structure 10 &0.805	&0.714	&0.834	&1.016\\
Experimental &\textbf{0.000}	&\textbf{0.000}	&\textbf{0.000}	&\textbf{0.000}\\
    \bottomrule
    \end{tabular}
    \label{table:charge_structures}
\end{table*}

One possible solution to this challenge is to explicitly provide information about the charge states of atoms during both training and structural optimization. By incorporating charge-state information into the MLIP representation, the model can account for configurational charge ordering prior to structural optimization, an important aspect of the optimization process, as demonstrated by the DFT results discussed above.

Following the approach of temporarily replacing \ce{Fe^{3+}} with \ce{Ga^{3+}} atoms, we identify all \ce{Fe^{3+}} atoms for each structure in the training dataset, using the magnetic moment as an indicator. During the MLIP training, we then explicitly encode the oxidation state of each \ce{Fe} atom by assigning separate encodings for \ce{Fe^{3+}} and \ce{Fe^{2+}}. Using this encoding-based approach, we retrain CHGNet, MACE, and MACELES such that each model can explicitly distinguish between \ce{Fe^{3+}} and \ce{Fe^{2+}} atomic sites. This enables the models to correctly identify the \ce{Fe^{3+}}/\ce{Fe^{2+}} arrangements and thereby account for the configurational charge ordering.


The improvement becomes evident when the charge-ordering-encoded MLIPs are used to perform structural optimization of the eleven structures in \cref{fig:GA_structures}, resulting in the relative energies shown in \cref{table:charge_structures}, together with the corresponding DFT values. All MLIP models now correctly identify the experimentally verified structure as the lowest-energy configuration. Furthermore, they reproduce the same ordering as obtained from DFT, correctly recognizing Structure 2 and Structure 1 as the second- and third-lowest-energy structures, respectively. Moreover, by incorporating configurational charge ordering into the MLIP representations, the models are able to reproduce the DFT configurational charge ordering distribution in \cref{fig:electronic_entropy}. As illustrated in \cref{app:electronic_entropy_with_MLIPs}, the MLIPs correctly identify the experimentally observed \ce{Fe^{2+}}/\ce{Fe^{3+}} spatial arrangement as the most energetically favorable configuration with overall low energy errors. These results demonstrate that encoding configurational charge ordering directly into the MLIP representation enables the models to recover the correct local chemical environments during structural optimization.

\begin{figure}
    \centering
    \includegraphics[width=1.0\textwidth]{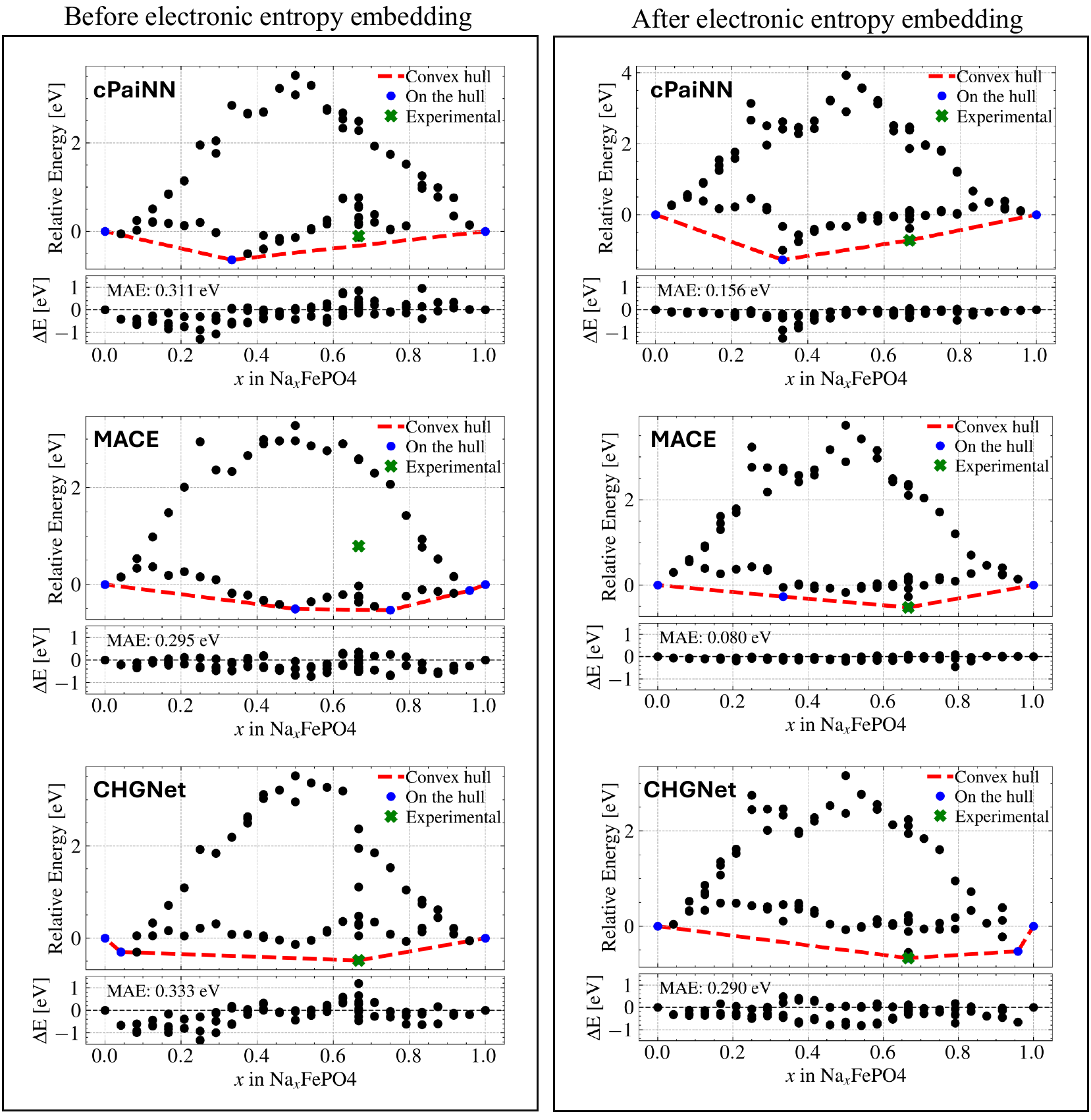}
    \caption{Convex hulls constructed using MLIPs before and after encoding configurational charge ordering, based on the DFT-optimized structures shown in \cref{fig:GA_result}c). Residuals are included to illustrate the differences between the MLIP-predicted and DFT relative energies.}
    \label{fig:convexhull}
\end{figure}

Having verified that the configurational charge ordering encoded MLIPs can correctly perform structural optimization and recover the energetically favorable local chemical environments, we return to the original objectives: constructing the convex hull for \ce{NaFePO4} and reproducing both the experimental results as well as the DFT-optimized result shown in \cref{fig:GA_result}c), which identifies stable configurations for \ce{NaFePO4}, \ce{FePO4}, and \ce{Na_{0.66}FePO4} as verified experimentally \cite{kyoto,kyoto_cif}. By optimizing all structures in \cref{fig:GA_result} using the MLIPs both before and after the charge-ordering-encoding, MLIP-based convex hulls can be constructed, as shown in \cref{fig:convexhull}. A clear observation across all MLIPs is that, after the charge-ordering-encoding, the experimentally verified stable configuration lies on the convex hull. In contrast, both MACE and MACELES fail to identify this configuration as stable before the charge-ordering-encoding. Moreover, the prediction errors for all MLIPs decrease significantly when configurational charge ordering is included, as reflected by the reduction in the mean absolute error (MAE) printed in the residuals plot. While CHGNet also shows improvement, the reduction in error is smaller compared with that observed for MACE and MACELES.

Comparing the predicted stable structures, those located on the convex hull, reveals that all models show improved agreement with the DFT convex hull shown in \cref{fig:GA_result}c) once the charge-ordering-encoding is included in the MLIP representation, which is clearly indicated by the residuals. However, MACE predicts a particularly stable structure at 33\% \ce{Na}-concentration that is not verified by DFT. Interestingly, the DFT calculations also indicate that a configuration at 33\% \ce{Na}-concentration lies relatively close to the convex hull, lending some support to the MACE prediction. In contrast, CHGNet identifies a stable structure at 95\% \ce{Na}-concentration that is not observed in the DFT results, making its predictions less consistent with the DFT convex hull. 

MACE and MACELES exhibit very similar predictions both before and after the explicit configurational charge ordering encoding. Although MACELES provides a more accurate overall energy prediction than MACE, the encoding results in only modest changes to the predicted convex hull. Nevertheless, MACELES produces the closest agreement with the DFT convex hull, as it is the only MLIP that predicts the experimentally observed set of stable phases without introducing additional spurious phases.

Overall, all MLIPs show substantial improvement once configurational charge ordering is encoded in the model representation, with MACELES providing the closest agreement with the DFT convex hull once charge state information is included.

To expand the main work to a more generalized example case, the charge-ordering-encoding is introduced to the entire polyanion cathode materials dataset, containing four different transition-metal ions: Fe, Mn, Co, and Ni. Two MACE models were trained, one without the charge-ordering-encoding and one with the charge-ordering-encoding, as detailed in \cref{sec:training_dataset}, with training curves shown in \cref{app:train_gen_MLIP}. To test the two models, the structure optimization dataset from Ref.~\cite{cPaiNN} is utilized, which contains structures with two to four transition-metal ions co-doped into supercells and unit cells of the same four polyanion cathode frameworks (olivine and maricite \ce{NaTMPO4}-type, \ce{Na2TMSiO4}, and \ce{Na_{2.56}TM_{1.72}(SO4)3}) with TM $\in$ {Fe, Mn, Co, Ni}. Because these multi-TM-ion doped structures were not included in the training set used in the present work, this dataset provides an out-of-distribution test of whether the charge-ordering encoding generalizes beyond the compositions and doping patterns seen during training. The MACE models optimize the structures starting from their pre-optimized state, and the resulting structures are compared to the corresponding DFT-optimized structures. Considering \cref{table:cell_relax}, we observe that the charge-encoded MACE model shows a reduced MAE across all metrics relative to the baseline model, with the energy MAE decreasing from 8.202 to 4.836 meV/atom. While this reduction in energy error is notable, the clearest improvement is seen in the structural metrics, where the mean absolute percentage error (MAPE) for the volume and the lattice parameters is substantially reduced. This indicates that, beyond improving energetic accuracy, explicitly encoding charge-ordering information has a pronounced effect on the structural fidelity of the relaxed geometries, suggesting that the model's ability to resolve the correct local charge environment during optimization is directly tied to recovering accurate lattice geometry, consistent with the mechanism identified for the Fe-only NaFePO$_4$ benchmark discussed above.
\begin{table*}[t!]
\centering
    \caption{MAE between DFT- and MACE-predicted total energy, and MAPE between DFT- and MACE-predicted volume (V), lattice parameters, and angles, for structural optimization results using MACE models trained on the entire polyanion sodium cathode materials dataset, with and without charge-ordering-encoding.}
    \footnotesize
    \begin{tabular}{ccccccccc}
    \br
    & E [meV/atom] & V [Å$^3$] & $a$ [Å] & $b$ [Å] & $c$ [Å] & $\alpha$ [$^\circ$] & $\beta$ [$^\circ$] & $\gamma$ [$^\circ$] \\ 
    \mr
    Without charge-ordering-encoding & 8.202 & 2.289 & 1.873 & 0.879 & 0.315 & 0.964 & 0.457 & 0.736 \\
    With charge-ordering-encoding & 4.836 & 0.786 & 0.333 & 0.785 & 0.266 & 0.925 & 0.155 & 0.544 \\ 
    \br
    \end{tabular}
    \label{table:cell_relax}
\end{table*}

\section{Conclusion}
In this work, we investigated the role of configurational charge ordering in machine learning interatomic potential (MLIP) simulations of charge-disordered battery cathode materials. Using \ce{NaFePO4} as a model system, we demonstrated that conventional MLIPs struggle to correctly predict the thermodynamic stability of mixed-valence materials during structural optimization. Specifically, we showed that MLIP-driven structural optimization frequently leads to incorrect \ce{Fe^{2+}}/\ce{Fe^{3+}} charge assignments, resulting in unfavorable local chemical environments and incorrect energy ordering of candidate structures.

Through systematic analysis of magnetic moments and charge ordering during structural optimization, we found that DFT largely determines the oxidation states of transition-metal atoms early in the optimization process. In contrast, conventional MLIPs fail to resolve the charge-state differentiation at this stage, which propagates through the structural optimization and ultimately leads to incorrect convex-hull predictions. Importantly, we demonstrated that MLIPs are capable of identifying the correct charge distribution when the correct local chemical environment is provided, indicating that the primary limitation lies in capturing the configurational charge ordering during structural optimization.

To address this limitation, we introduced an approach for encoding configurational charge ordering directly into MLIP representations by explicitly encoding the \ce{Fe^{2+}} and \ce{Fe^{3+}} oxidation states during model training. Using this strategy, we retrained the MLIPs, CHGNet, MACE, and MACELES, which demonstrated that the resulting charge-ordering-encoded MLIPs successfully identify the experimentally observed stable configurations and reproduce the DFT energy ordering across sodium concentrations. 

Beyond the\ce{NaFePO4} benchmark, we demonstrated that this charge-ordering-encoding strategy generalizes to a substantially broader materials space, by applying the same approach to a MACE model trained across the entire polyanion sodium cathode materials dataset, spanning four crystal frameworks and four transition-metal ions (\ce{Fe}, \ce{Mn}, \ce{Co}, \ce{Ni}), yielded consistent improvements in both energetic and structural accuracy when tested on out-of-distribution structures containing multiple co-doped transition-metal species, indicating the benefits of explicit charge-state encoding on a larger scale.

Beyond identifying the charge ordering, the encoding strategy introduced here offers an additional practical advantage. Because the charge state is fixed through the atomic species label itself, a specific charge-ordering arrangement can be imposed and held fixed throughout the entire structural optimization. Unconstrained DFT struggles to guarantee this, as charge can redistribute between transition-metal sites as atomic positions evolve. This makes the charge-ordering-encoded MLIP useful as a pre-optimizer, allowing the energetic consequences of a specific configuration to be isolated and studied without the confound of charge delocalization that complicates the equivalent DFT calculation. Our results demonstrate that incorporating configurational charge ordering into the MLIP representation is essential for modeling charge-disordered materials. More broadly, this work highlights the importance of including electronic degrees of freedom in MLIPs when modeling materials whose thermodynamic stability depends strongly on charge ordering. We anticipate that similar approaches will be necessary for reliable MLIP simulations of a wide range of transition-metal compounds, including battery cathodes, catalytic materials, and strongly correlated oxides.

One limitation of this work is that the number of degrees of freedom increases, as the system now includes not only the positions of the \ce{Na} atoms but also the arrangement of \ce{Fe^{3+}} and \ce{Fe^{2+}} ions. This increases the computational cost, since the optimization algorithms must determine both the optimal \ce{Na} positions and the optimal \ce{Fe^{3+}}/\ce{Fe^{2+}} charge ordering. One possible solution is to employ a dedicated optimization algorithm to determine the optimal arrangement of \ce{Fe^{3+}} and \ce{Fe^{2+}} ions for each \ce{NaFePO4} configuration (e.g., an Ewald summation scheme). Alternatively, the configurational charge ordering could be incorporated indirectly into the MLIP representation, such that the model learns the charge distribution without requiring the explicit specification of \ce{Fe^{3+}}/\ce{Fe^{2+}} arrangements for every configuration. Exploring such approaches represents an interesting direction for future work.

\section{Methods}
%
\subsection{Density Functional Theory}
All DFT calculations were performed using the Vienna Ab initio Simulation Package (VASP, version 6.4)\cite{vasp}. The Perdew–Burke–Ernzerhof (PBE) exchange–correlation functional\cite{PBE}, supplemented with Hubbard \textit{U} corrections, was used throughout to mitigate the electronic self-interaction error, which can strongly affect systems exhibiting localized \textit{d}-orbital electrons. The applied value of \ce{U_{Fe}} = \qty{5.3}{\electronvolt} was adopted from experimentally validated reference structures in the Materials Project database\cite{MP_U_value, MP_database}. A plane-wave energy cutoff of \qty{520}{\electronvolt} was employed for all calculations, together with Gaussian smearing using a width of \qty{0.01}{\electronvolt}. Electronic self-consistency was converged to within \qty{1e-5}{\electronvolt}, while ionic relaxation was considered complete when all atomic forces fell below \qty{0.03}{\electronvolt/\angstrom}. Spin polarization was included in all calculations. Brillouin zone integrations were carried out using $\Gamma$-centered \textit{k}-point grids with a fixed resolution of $(3 \times 1 \times 3)$ applied consistently to all structures. All computational parameters follow the protocol established in the polyanionic sodium cathode materials dataset\cite{Polyanion_dataset}.

\subsection{Benchmark datasets}
\label{sec:benchmark_datasets}
For the genetic algorithm optimization, an orthorhombic olivine \ce{NaFePO4} cathode material, adapted from Ref.~\cite{Simon}, was expanded to a $(1 \times 3 \times 2)$ supercell, resulting in 24 \ce{Na} sites and enabling the construction of structures at 66\% sodium concentration. The DFT-optimized dataset used to construct the convex hull in \cref{fig:GA_result}c) consists of 146 structures, since one structure at 76\% sodium concentration did not converge. The benchmark dataset includes the experimentally verified configuration highlighted in green. This experimental structure obtained by mapping the \ce{Na} ordering of the originally monoclinic \cite{kyoto_cif}, into the orthorhombic olivine cell used in this work. This dataset includes energies, forces, magnetic moments, and charge analyses, and serves as the reference for training and benchmarking the MLIPs considered in this study.

An additional dataset is constructed to investigate the contribution of configurational charge ordering during DFT optimization. This dataset is obtained by considering, all possible spatial arrangements of \ce{Fe^{3+}}/\ce{Fe^{2+}} for the experimental verified configuration. This corresponds to a total of 735,472 possible configurations. As it is computationally infeasible to evaluate all configurations, a random subset of 171 structures was selected for analysis. To ensure a specific spatial arrangement of \ce{Fe^{3+}}/\ce{Fe^{2+}}, a two-step method is deployed. First, a geometry optimization is performed in which all \ce{Fe} ions intended to be \ce{Fe^{3+}} are substituted by \ce{Ga} ions. This exploits the fact that \ce{Ga} preferentially adopts a stable +3 oxidation state (with +2 being energetically unfavorable) and has an ionic radius similar to that of \ce{Fe}. Second, the optimized structure from the first step is used as a starting point, and the \ce{Ga} ions are replaced back with \ce{Fe}. A subsequent geometry optimization is then carried out with initialized magnetic moments of approximately 4$\mu_B$ for \ce{Fe^{2+}} and 5$\mu_B$ for \ce{Fe^{3+}}. This two-step procedure, developed in Ref.~\cite{Ga_method}, ensures that oxidation occurs at predefined positions while preserving ferromagnetic ordering, as demonstrated in Ref.~\cite{hao2015coexistence}.

The benchmark dataset and the configurational charge ordering dataset can be accessed at \url{https://doi.org/10.11583/DTU.31812499}.

\subsection{Training dataset}
\label{sec:training_dataset}
Initially, a MACE (v0.3.13) \cite{batatia2022mace} model trained on the polyanion sodium cathode materials dataset \cite{Polyanion_dataset} was used to drive the genetic algorithm optimization and construct the composition--energy convex hull shown in \cref{fig:GA_result}b. The DFT convex hull in \cref{fig:GA_result}c was subsequently obtained by optimizing the low-energy structures identified from this MLIP-driven search. The trained model and the corresponding training, validation, and test splits are available in Ref.~\citenum{Na_dataset_test}.

To investigate the role of charge ordering, a new training dataset containing only \ce{Fe}-based sodium polyanion cathode materials was constructed by removing all compounds in which \ce{Fe} was not the sole transition-metal species.  This dataset consist of 28,455 crystal structures spanning a range of relevant cathode materials with varying sodium concentrations, including olivine and maricite \ce{NaFePO4}, \ce{Na2FeSiO4}, and \ce{Na_{2.56}Fe_{1.72}(SO4)3}. 

To ensure a clear identification of the redox-active \ce{Fe^{2+}} and \ce{Fe^{3+}} species, configurations corresponding to transition states encountered during structural optimization and molecular dynamics (MD) simulations were systematically removed from the training data. These intermediate geometries often lack well-defined electronic states and can obscure the energetic separation between different charge configurations. To ensure an unambiguous assignment of oxidation states, structures containing \ce{Fe} atoms with magnetic moments outside the intervals of 3.4--3.8~$\mu_\mathrm{B}$ and 4.2--4.4~$\mu_\mathrm{B}$ were discarded. The remaining Fe atoms were assigned as \ce{Fe^{2+}} and \ce{Fe^{3+}} according to their magnetic moments, and the corresponding Bader charges and atomic magnetic moments were updated to reflect these oxidation states. After this filtering the training dataset consist of a total of 24,766 structures.

Two versions of this dataset were subsequently used for retraining the MLIP models. In the first, the conventional representation was retained, where all iron atoms were treated as the same atomic species (\ce{Fe}). In the second, the configurational charge ordering was explicitly encoded into the atomic representation by replacing each \ce{Fe^{3+}} atom with a surrogate atomic species, \ce{Ga^{3+}}, while \ce{Fe^{2+}} atoms retained the \ce{Fe} label. 

For the generalized charge-ordering MLIP model, the whole polyanion sodium cathode materials dataset \cite{Polyanion_dataset} was filtered to ensure that any ambiguous assignment of oxidation states was discarded. The initial dataset contains 298,144 structures. After removing structures containing \ce{Fe} atoms with magnetic moments outside the intervals of 3.4–3.8 $\mu_\mathrm{B}$ and 4.2–4.4 $\mu_\mathrm{B}$, the dataset was reduced to 293,491 structures. Next, structures containing \ce{Ni} atoms with magnetic moments outside the intervals of 1.7–1.9 $\mu_\mathrm{B}$ and 2.0–2.5 $\mu_\mathrm{B}$ were discarded, resulting in 287,772 structures. Next, structures containing \ce{Mn} atoms with magnetic moments outside the intervals of 3.7–4.2 $\mu_\mathrm{B}$ and 4.5–5.2 $\mu_\mathrm{B}$ were discarded, resulting in 243,643 structures. Next, structures containing \ce{Co} atoms with magnetic moments outside the intervals of 2.6–2.84 $\mu_\mathrm{B}$ and 2.84–3.3 $\mu_\mathrm{B}$ were discarded, resulting in 226,443 structures. Assigning the charge state to each individual atom, resulted in materials with non-zero charge; after removing these non-charge-neutral structures, 172,696 structures remained. As before, two versions of this dataset were subsequently used for retraining the MACE model. In the first, the conventional representation was retained. In the second, oxidation states were explicitly encoded by representing \ce{Mn^{3+}} with \ce{In}, \ce{Co^{3+}} with \ce{Al}, and \ce{Ni^{3+}} with \ce{Ti}, enabling a generalized MACE model to be trained that accounts for optimized structures across different transition-metal ions in the polyanion sodium cathode material phase space. For the test set, the structure optimization dataset from Ref.~\cite{cPaiNN} is used, which contains structures with two to four transition-metal ions co-doped into supercells and unit cells of the same four polyanion cathode frameworks (olivine and maricite \ce{NaTMPO4}-type, \ce{Na2TMSiO4}, and \ce{Na_{2.56}TM_{1.72}(SO4)3}) with TM $\in$ {Fe, Mn, Co, Ni}. The same replacement is used for the training dataset and structures identified without a total charge state of zero was removed from the test set.

The training dataset with training and validation splits are available at \url{https://doi.org/10.11583/DTU.31812499}.

\subsection{Machine Learning Interatomic Potentials}
CHGNet (v0.4.1)\cite{deng2023chgnet}, MACE (v0.3.13)\cite{batatia2022mace}, and MACELES (v0.3.14)\cite{Kim2026} were all deployed in this work. 

For MACE and MACELES, the parameters of the MACE-MP-0 large universal model \cite{batatia2023foundationmace} were adopted without modification. CHGNet models employed the same architecture and hyperparameters as the publicly available universal CHGNet potential. 

All models were trained from scratch to ensure that no external datasets influenced the training. They were all also trained on the datasplit of the training dataset, ensuring consistency. The training curves and the end result of training the three MLIPs with and without the explicit configurational charge ordering encoding is stated in \cref{app:training}.

All structural optimizations performed with MLIPs employed the LBFGSLineSearch optimizer together with the FrechetCellFilter for cell relaxation, both implemented in ASE \cite{ASE}. Convergence was achieved when the maximum force on any atom was below \SI{0.01}{\electronvolt/\angstrom}. In some cases, especially before adding the configurational charge ordering in the MLIP, some structures are not converged and thus excluded in \cref{fig:convexhull}.

The final trained models used for benchmarking in this study are available at \url{https://doi.org/10.11583/DTU.31812499}. 

\subsection{Genetic Algorithm}
The GA \cite{Lysgaard2013} implemented in the Atomic Simulation Environment (ASE, v3.26.0) \cite{ASE} was employed to efficiently sample low-energy disordered configurations of \ce{NaFePO4} across intermediate sodium concentrations. The large number of possible arrangements of \ce{Na^+} ions and vacancies over the crystallographic \ce{Na} Wyckoff sites makes exhaustive enumeration computationally impractical. The GA therefore provides an effective global optimization strategy to explore this configurational space using energies predicted by the MLIPs.

Each GA run was initialized with a population of symmetrically distinct structures, generated by randomly distributing \ce{Na} ions over the available sites. The full range of Na concentrations was sampled to ensure sufficient diversity in the initial population. The fitness of each individual was evaluated using the relative energy
\begin{equation}
E_{\mathrm{rel}}(x) = E(\ce{Na}_x\ce{FePO4}) - x\cdot E(\ce{NaFePO4}) - (1 - x)\cdot E(\ce{FePO4}),
\label{eq:rel_energy}
\end{equation}
where $x \in [0,1]$ denotes the sodium concentration. These relative energies were used to construct the composition–energy convex hull, with energies expressed in \si{\milli\electronvolt} per formula unit (f.u.) as predicted by the MLIPs.

To ensure balanced sampling across \ce{Na} concentrations, the fitness was ranked within each composition, and structures with equal rank were assigned identical effective fitness during selection. Standard GA operations—selection, crossover, and mutation—were then applied to evolve the population toward lower energies. Roulette-wheel selection was used to preferentially retain structures with higher fitness, while crossover operations combined \ce{Na} distributions from parent structures to generate new candidate configurations. Mutations were introduced by randomly swapping \ce{Na} ions with vacancies or relocating ions between sites, thereby maintaining diversity in the population and reducing the risk of premature convergence.

The algorithm was iterated for multiple generations until convergence was reached, defined as the absence of further improvement in the population energy.



\textbf{Acknowledgements}
A.B., J.M.G.L., and M.H.P. acknowledge support from the Det Frie Forskningsråd under Project ``Data-driven quest for TWh scalable \ce{Na}-ion battery (TeraBatt)'' (Ref. Number 2035-00232B). 

\textbf{Data and code accessibility}
The code used for this work is available at \url{https://github.com/dtu-energy/cPaiNN}. 

The training dataset, the trained MLIPs, and the benchmark dataset utilized in this work are available at: \url{https://doi.org/10.11583/DTU.31812499}

\bibliography{refs}
\newpage

\begin{appendices} 
\section{Training the machine learning interatomic potentials}
\label{app:training}

All MLIPs were trained using the identical training, validation, and test splits described in Section~\ref{sec:training_dataset}. For each architecture, two independent models were trained: (i) a baseline model using the conventional atomic representation and (ii) a model employing the explicit configurational charge ordering encoding, in which \ce{Fe^{3+}} ions were represented by a surrogate atomic species while \ce{Fe^{2+}} ions retained the \ce{Fe} label.

\begin{figure}[htb]
    \centering
    \includegraphics[width=\linewidth]{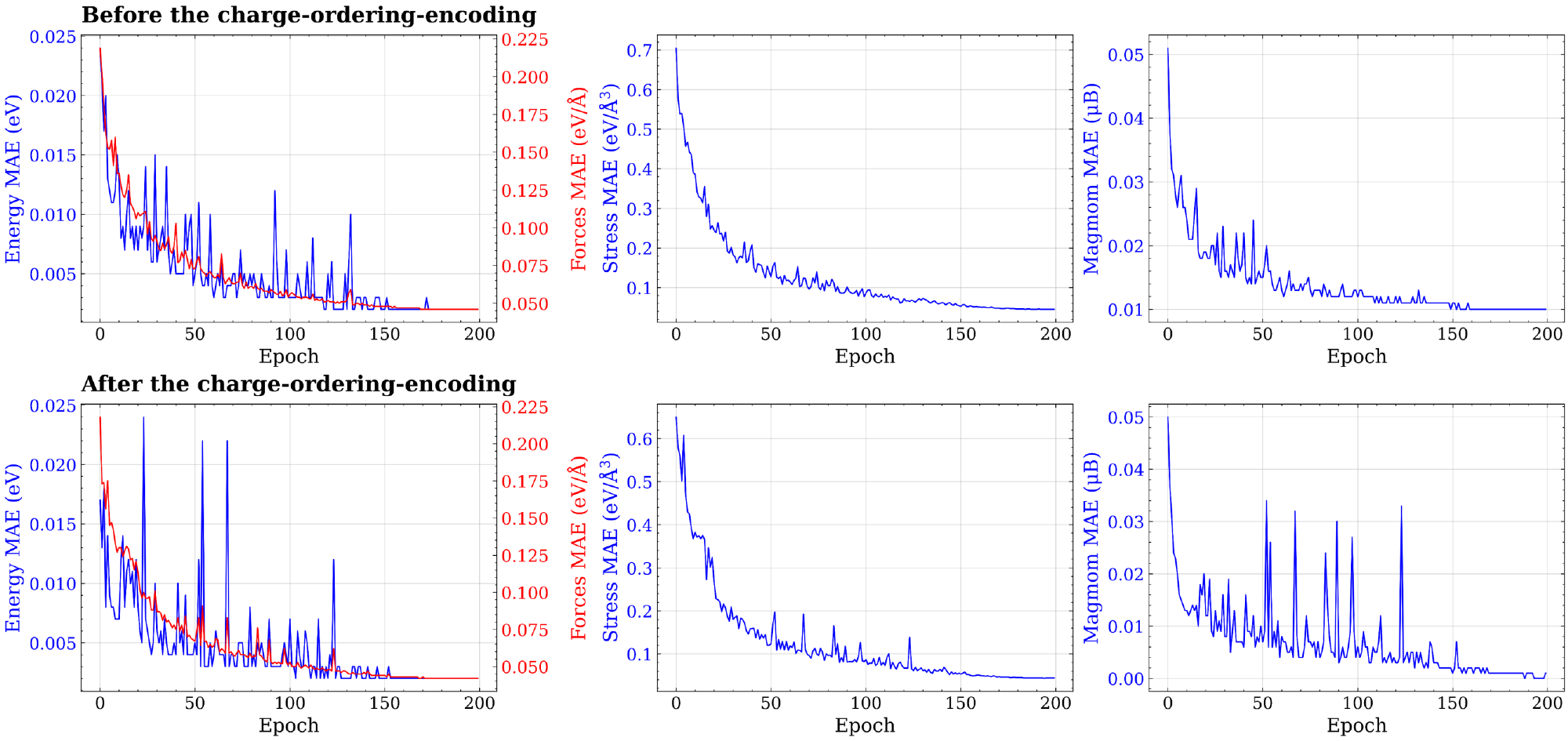}
    \caption{Validation learning curves for the CHGNet models trained before and after including configurational charge ordering. The explicit encoding leads to lower validation errors for all predicted properties, particularly the energy, and magnetic moments.}
    \label{fig:CHGNet_train}
\end{figure}

\begin{figure}[htb]
    \centering
    \includegraphics[width=0.8\linewidth]{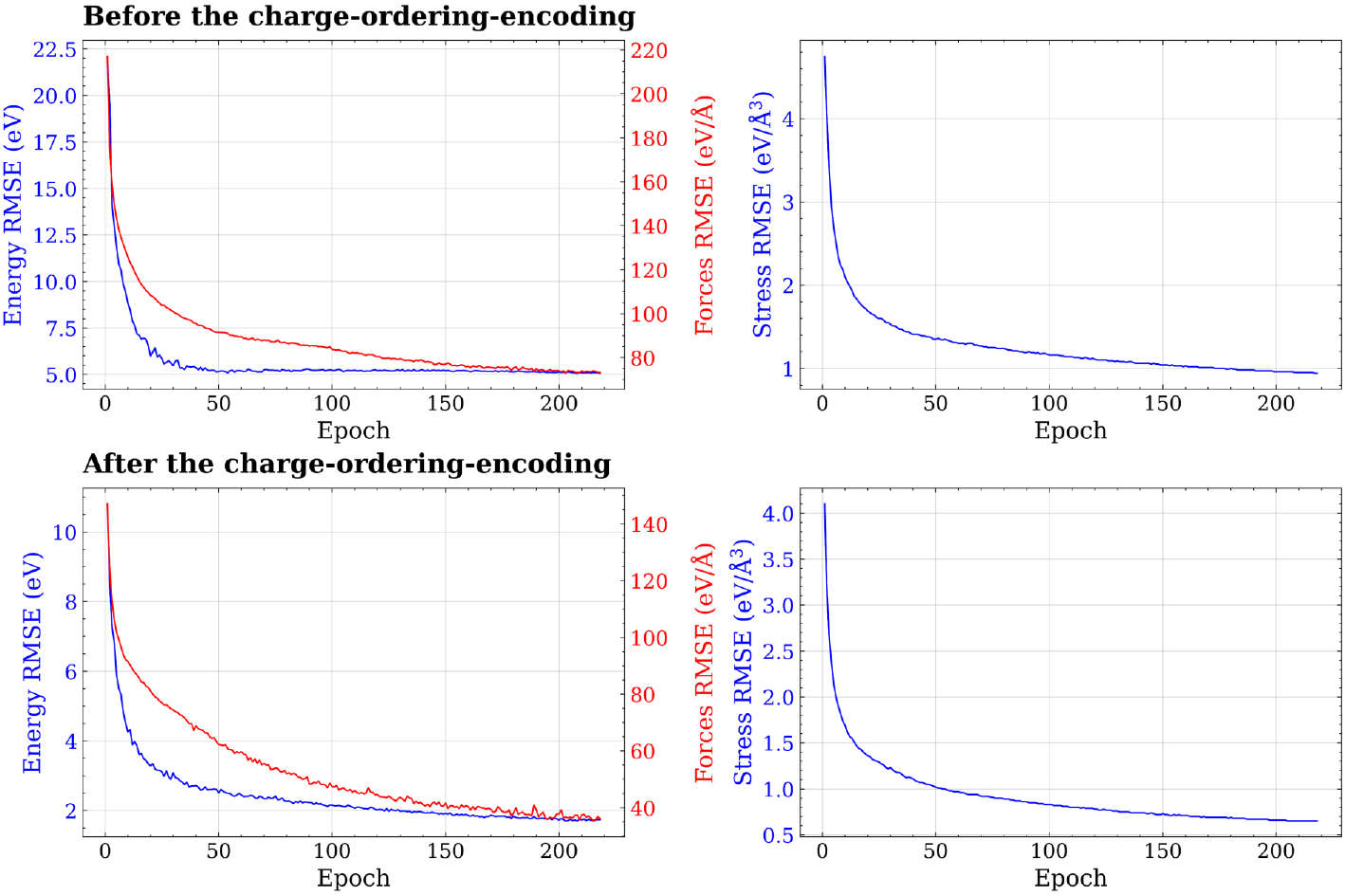}
    \caption{Validation learning curves for the MACE models trained before and after including configurational charge ordering. The explicit encoding leads to lower validation errors for all predicted properties, particularly the energy and forces}
    \label{fig:MACE_train}
\end{figure}

\begin{figure}[htb]
    \centering
    \includegraphics[width=0.8\linewidth]{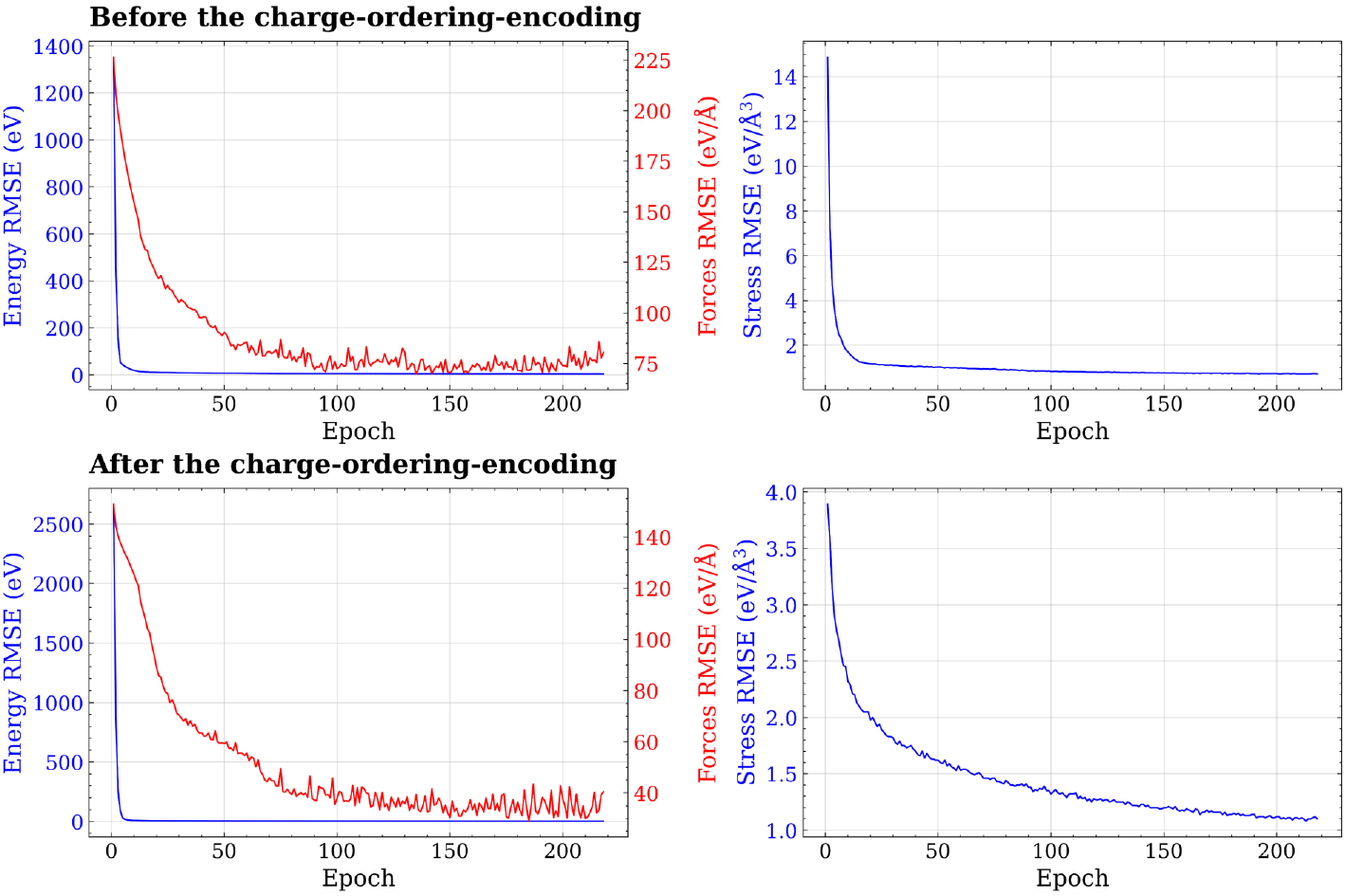}
    \caption{Validation learning curves for the MACELES models trained before and after including configurational charge ordering. Similar to MACE, the explicit encoding improves the convergence of the energy and force predictions, demonstrating that incorporating charge-state information facilitates learning in mixed-valence materials.}
    \label{fig:MACELES_train}
\end{figure}

The resulting learning curves are shown for CHGNet (\cref{fig:CHGNet_train}), MACE (\cref{fig:MACE_train}), and MACELES (\cref{fig:MACELES_train}). Across all three architectures, explicitly encoding the charge state consistently reduces the training error compared to the conventional representation. For CHGNet, the final validation MAE decreased from 46.0 to 42.0 meV~\AA$^{-1}$ for the forces, from 45.0 to 44.0 meV~\AA$^{-3}$ for the stress, and from 0.010 to 0.0010~$\mu_\mathrm{B}$ for the magnetic moment, while maintaining an energy MAE of 2.0 meV atom$^{-1}$. 

For the MACE and MACELES models, the explicit charge-ordering-encoding likewise improved the final validation RMSE. MACE reduced the energy RMSE from 5.06 to 1.74 meV atom$^{-1}$, the force RMSE from 73.19 to 36.19 meV~\AA$^{-1}$, and the stress RMSE from 0.94 to 0.65 eV~\AA$^{-3}$. MACELES exhibited similar behavior, with the energy RMSE decreasing from 3.62 to 2.87 meV atom$^{-1}$ and the force RMSE decreasing from 80.51 to 40.51 meV~\AA$^{-1}$, although the stress RMSE increased slightly from 0.72 to 1.10 eV~\AA$^{-3}$. Overall, the reduction in prediction errors across all three architectures indicates that explicitly distinguishing between \ce{Fe^{2+}} and \ce{Fe^{3+}} substantially simplifies the learning problem and enables the MLIPs to learn a more accurate representation of the potential energy surface for mixed-valence systems.

\section{Atomic charge ordering of machine learning interatomic potentials}
\label{app:charge_singlepoint}
To verify that the local chemical environments obtained from MLIP structural optimizations in \cref{fig:MLIP_scatter} correspond to alternative less energetically favorable configurations rather than incorrect solutions, single-point DFT calculations were performed on the MLIP-optimized experimental structures and compared with the fully DFT-optimized reference structure.

\begin{figure}[b!]
    \centering
    \includegraphics[width=0.75\textwidth]{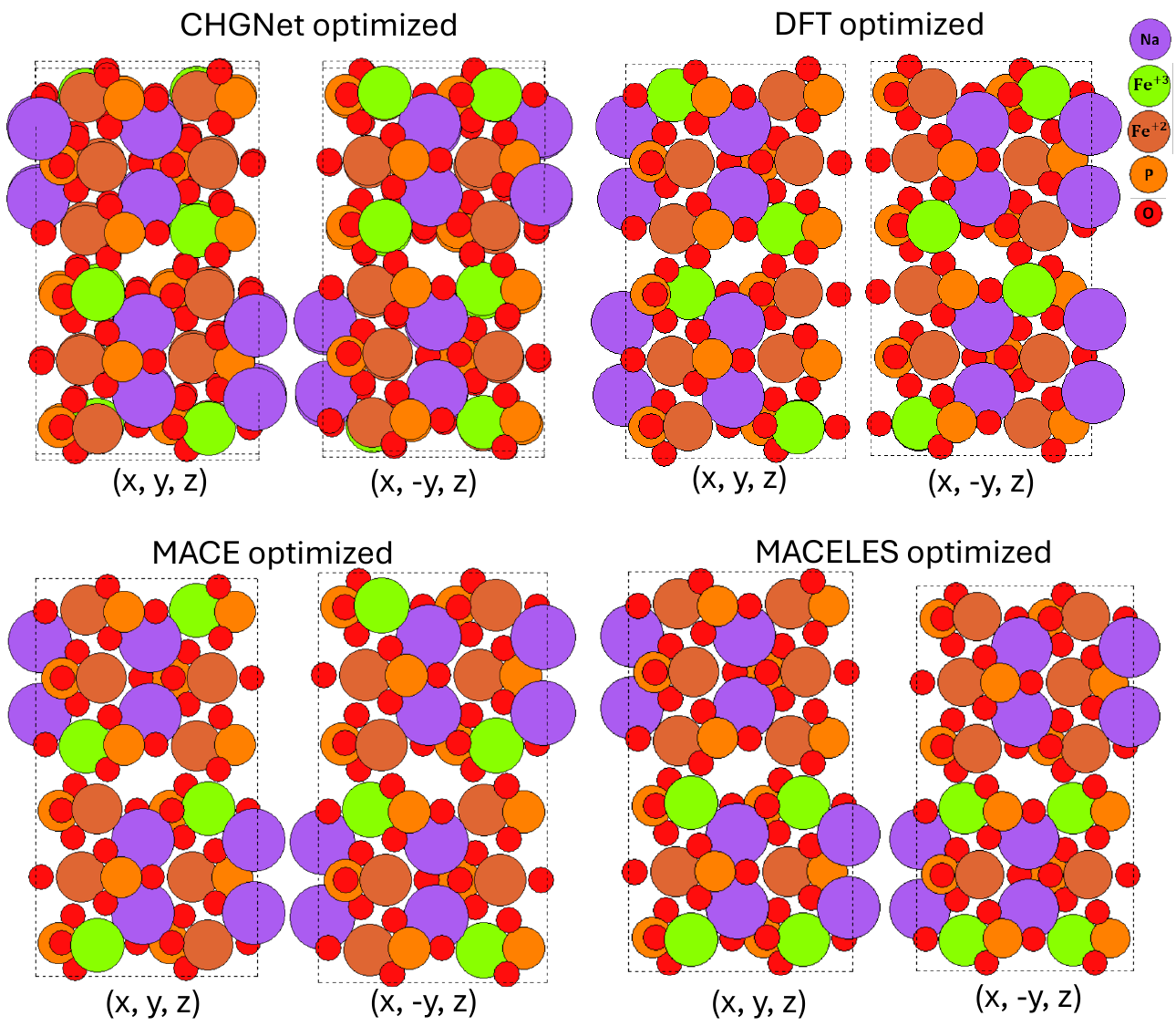}
    \caption{Single-point DFT calculation of the MLIP-optimized experimentally verified structure of \ce{Na_{0.66}FePO4}, illustrating the charge ordering of \ce{Fe^{3+}} and \ce{Fe^{2+}}.The charge ordering predicted by CHGNet follows the found charge ordering of the single-point DFT calculations.}
    \label{fig:charge_structures}
\end{figure}

As shown in \cref{fig:charge_structures}, the spatial distributions of \ce{Fe^{3+}} and \ce{Fe^{2+}} differ between the MLIP-optimized and DFT-optimized structures. The CHGNet produces a charge ordering that is largely similar to the DFT result, although a small number of \ce{Fe} atoms are misassigned. In contrast, MACELES separates the \ce{Fe^{3+}} and \ce{Fe^{2+}} into a two phase regime, which is in disagreement with the DFT predictions. MACE predicts identical charge-ordering patterns that appear as mirrored configurations relative to the DFT-optimized structure. These observations further support the conclusion drawn in \cref{fig:MLIP_scatter}, namely that MLIPs converge to alternative, less energetically favorable local chemical environments during structural optimization.

Despite these differences in charge ordering, the magnetic moments predicted by CHGNet are consistent with those obtained from single-point DFT calculations. Similarly, the predicted energies are in close agreement with single-point DFT calculations, indicating that MLIPs are capable of accurately capturing both the configurational charge ordering and the energetics when the local chemical environment is fixed, but they fail to identify the correct global minimum during structural optimization.




\section{Machine learning interatomic potentials performance on configurational charge ordering}
\label{app:electronic_entropy_with_MLIPs}
To demonstrate both the importance of configurational charge ordering for MLIPs and the capability of charge-ordering-encoded models, we apply MLIP-based structural optimization to the configurational charge ordering dataset introduced in \cref{fig:electronic_entropy}. This dataset, which was originally used to illustrate the role of configurational charge ordering during DFT optimization, includes the experimentally verified structure as a reference.

\begin{figure}
    \centering
    \includegraphics[width=1\textwidth]{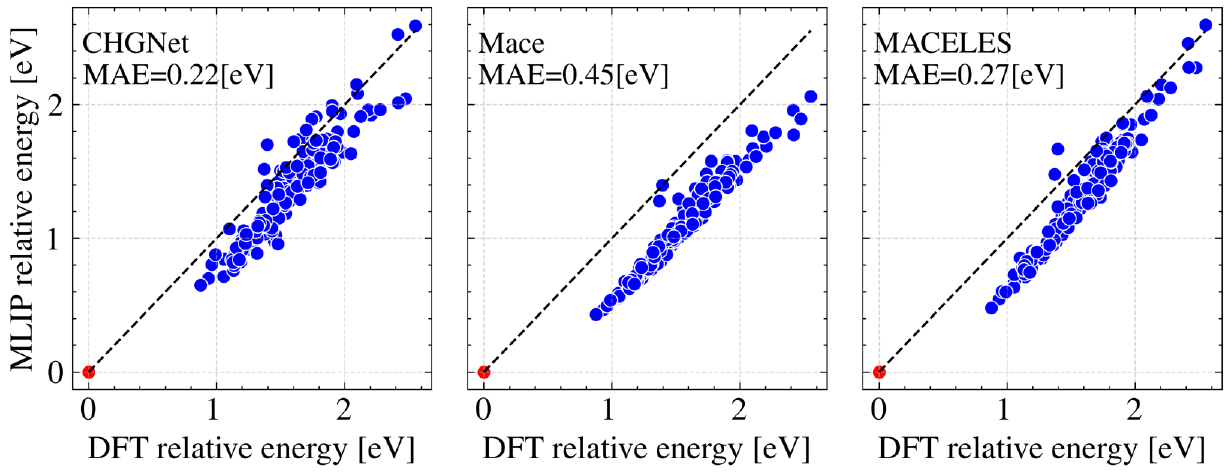}
    \caption{Relative energies of the charge-ordering-encoded MLIP-optimized structures for all 171 configurations shown in \cref{fig:electronic_entropy}, referenced to the charge-ordering-encoded MLIP-optimized energy of the experimentally verified structure (highlighted in red). The MLIP relative energies are plotted against the DFT relative energies from \cref{fig:electronic_entropy}}
    \label{fig:electronic_entropy_with_MLIPs}
\end{figure}

As shown in \cref{fig:electronic_entropy_with_MLIPs}, the charge-ordering-encoded MLIPs are able to clearly distinguish between different \ce{Fe^{2+}}/\ce{Fe^{3+}} arrangements, assigning higher energies to less favorable charge-ordering configurations relative to the experimentally optimized structure. This behavior is consistent with the DFT results and demonstrates that the charge-encoded models correctly capture the energetic landscape associated with configurational charge ordering.

\section{Training the generalized machine learning interatomic potential to consider multiple transition metal ions.}
\label{app:train_gen_MLIP}
Like for the other MLIPs two independent models were trained: (i) a baseline model using the conventional atomic representation and (ii) a model employing the explicit configurational charge ordering encoding, in which all transition metal ions with charge state +3 (\ce{TM^{3+}}) ions were represented by a surrogate atomic species while \ce{TM^{2+}} ions retains their original label. The train and validation split was kept consistent across models. The four transition-metal ions were explicitly encoded by representing \ce{Mn^{3+}} with \ce{In}, \ce{Co^{3+}} with \ce{Al}, and \ce{Ni^{3+}} with \ce{Ti}..

The resulting learning curves indicate a clear lowering of the energy, forces and stress when the charge-ordering-encoding is utilized. The energy RMSE is reduced from 6.34 to 1.35 meV atom$^{-1}$, the force RMSE from 53.22 to 17.4 meV~\AA$^{-1}$, and the stress RMSE from 0.69 to 0.52 eV~\AA$^{-3}$. This indicates that explicitly distinguishing between the charge state of the transition metal ions substantially simplifies the learning problem and enables the MLIPs to learn a more accurate representation of the potential energy surface for mixed-valence systems.

\begin{figure}[htb]
    \centering
    \includegraphics[width=0.8\linewidth]{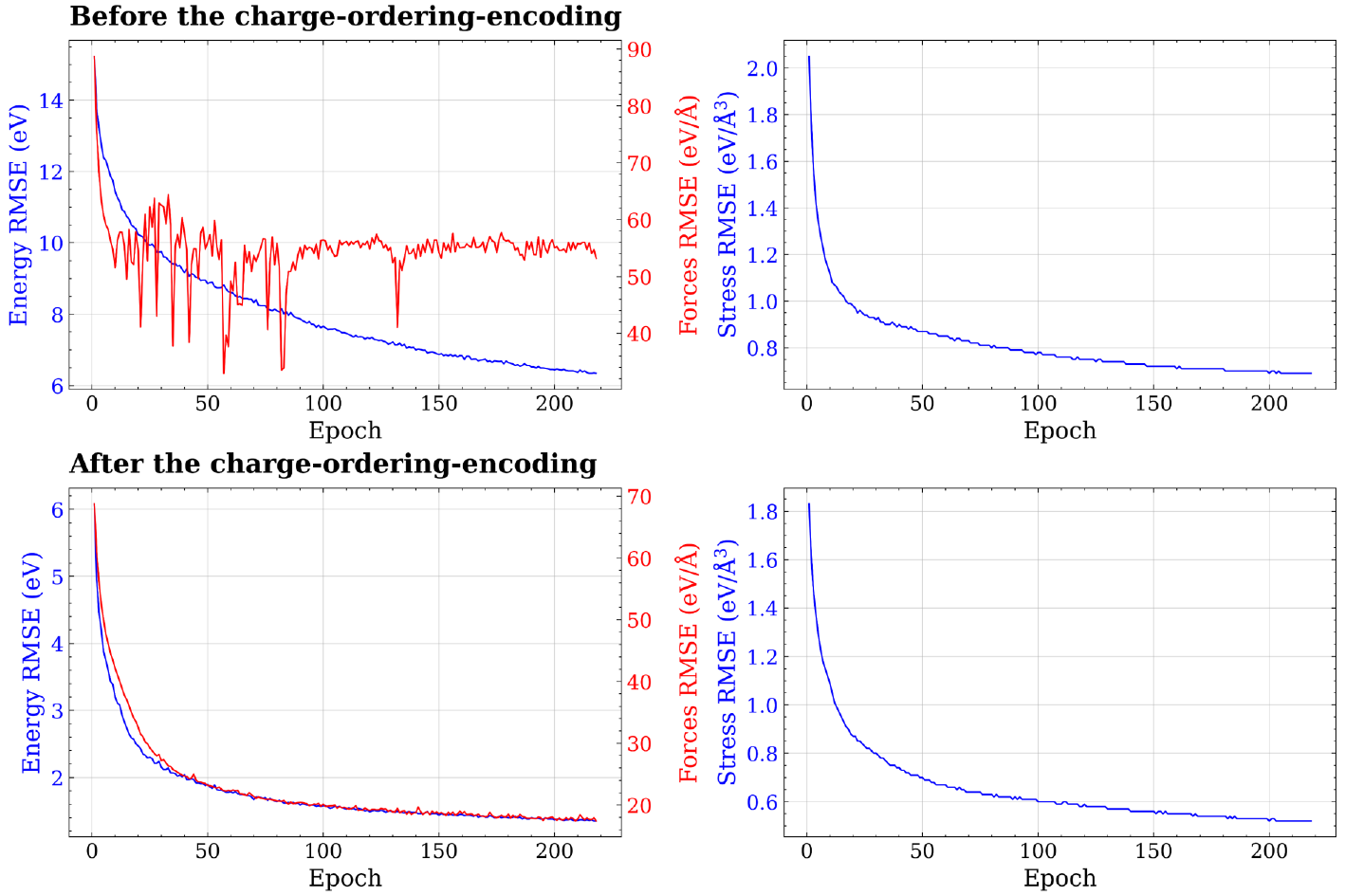}
    \caption{Validation learning curves for the MACE models trained before and after including configurational charge ordering on the entire polyanion sodium cathode materials dataset, including four different transition metal ions. The explicit encoding leads to lower validation errors for all predicted properties, particularly the energy and forces}
    \label{fig:MACE_polyanion_train}
\end{figure}

\end{appendices}

\end{document}